\documentclass[%
 reprint,
nofootinbib,
 amsmath,amssymb,
 aps,
]{revtex4-2}
\usepackage[export]{adjustbox}
\usepackage{graphicx}
\usepackage{dcolumn}
\usepackage{bm}

\begin{document}

\preprint{APS/123-QED}

\title{Multiple Series Representations of $N$-fold Mellin-Barnes Integrals}

\author{B. Ananthanarayan and Sumit Banik}
 \affiliation{Centre for High Energy Physics, Indian Institute of Science,
Bangalore-560012, Karnataka, India}

\author{Samuel Friot}
\affiliation{
 Universit\'e Paris-Saclay, CNRS/IN2P3, IJCLab, 91405 Orsay, France \\
 Univ Lyon, Univ Claude Bernard Lyon 1, CNRS/IN2P3,
 IP2I Lyon, UMR 5822, F-69622, Villeurbanne, France}

\author{Shayan Ghosh}
\affiliation{Helmholtz-Institut f\"ur Strahlen- und Kernphysik, D-53115 Bonn, Germany \\
Bethe Center for Theoretical Physics, Universit\"at Bonn, D-53115 Bonn, Germany}

\begin{abstract}
Mellin-Barnes (MB) integrals are  well-known objects appearing in
many branches of mathematics and physics, ranging from hypergeometric functions theory to 
quantum field theory, solid state physics, asymptotic theory, etc. Although MB integrals have been studied for more than one century, until now there is no systematic computational technique of the multiple series representations of $N$-fold MB integrals for $N>2$. Relying on a simple geometrical analysis based on conic hulls, we show here a solution to this important problem. Our method can be applied to resonant (\textit{i.e} logarithmic) and nonresonant cases and, depending on the form of the MB integrand, it gives rise to convergent series representations or diverging asymptotic ones. When convergent series are obtained the method also allows, in general, the determination of a single \textit{master series} for each series representation, which considerably simplifies convergence studies and/or numerical checks. We provide, along with this paper, a \textit{Mathematica} implementation of our technique with examples of applications. Among them, we present the first evaluation of the hexagon and double box conformal Feynman integrals with unit propagator powers.

\end{abstract}

\maketitle

\indent{\textit{Introduction}---} 
$N$-fold Mellin-Barnes (MB) integrals are defined as
\begin{align} \label{N_MB}
    &I (x_1,x_2,\cdots ,x_N) \nonumber \\
    & = \int\limits_{-i \infty}^{+i \infty} \frac{ d z_1}{2 \pi i} \cdots \int\limits_{-i \infty}^{+i \infty}\frac{ d z_N}{2 \pi i}\,\,  \frac{\prod\limits_{i=1}^{k} \Gamma^{a_i}(s_i ({\bf z})+g_i)}{\prod\limits_{j=1}^{l} \Gamma^{b_j}(t_j ({\bf z})+h_j)} x^{z_1}_{1} \cdots x^{z_N}_{N}
\end{align}
where $a_i , b_j, k, l$ and $N$ are positive integers (with $k\geq N$ after possible cancellations due to the denominator), ${\bf z}=(z_1,...,z_N)$ and where we have defined $s_i({\bf z}) \doteq {\bf e}_i\cdot{\bf z}$ and $t_j({\bf z}) \doteq {\bf f}_j\cdot{\bf z}$ for later purpose. The vectors ${\bf e}_i, {\bf f}_j$ and the scalars $g_i,h_j$ are reals while ${\bf x}=(x_1,...,x_N)$ can be complex, and the contours of integration, which avoid the poles of the gamma functions that belong to the numerator of the MB integrand, have to be specified. In the present work, we focus on the common situation where the set of poles of each of these gamma functions is not splitted in different subsets by the contours.

The importance of MB integrals cannot be overstated, as they appear in domains as diverse as hypergeometric functions theory \cite{KdF,Exton,Marichev}, electromagnetic wave propagation in turbulence \cite{Sasiela}, asymptotics \cite{Paris&Kaminsky}, quantum field theory (QFT) \cite{Smirnov:2012gma},  etc.
In QFT, which is of particular interest for the authors, an impressive array of publications of the last decades may be mentioned (see \cite{Smirnov:2012gma} for a complementary list). Early studies can be found in
\cite{Bjorken:1963zz,Trueman:1963zz,Usyukina1975}
, followed by classical works \cite{Boos:1990rg,Davydychev:1990jt,Davydychev:1990cq,Usyukina:1992jd,Usyukina:1993ch,Davydychev:1992mt,Berends_1994,Usyukina:1994iw,Smirnov:1999gc,Tausk:1999vh,Smirnov:2000vy,Smirnov:2001cm,Smirnov:2003vi,Heinrich:2004iq} highlighting the relevance of MB integrals in QFT. These motivated the automatization of some of the computational steps of the MB technique \cite{Czakon:2005rk,Anastasiou:2005cb,Gluza:2007rt,Smirnov:2009up,Ochman:2015fho}. 
Numerous applications were guided by the needs of particle physics phenomenology, \textit{e.g.} \cite{Friot:2005cu,Aguilar:2008qj,Czakon:2007wk,Smirnov:2009fh,Lee:2010cga,Greynat:2012ww,deRafael:2014gxa,Ananthanarayan:2016pos,Charles:2017snx,Ananthanarayan:2017qmx,Ananthanarayan:2018irl,Ananthanarayan:2020acj} but also by more formal motivations \cite{Bern:2005iz,Bern:2006vw,Bern:2006ew,Drummond:2006rz,Kalmykov:2008ofy,DelDuca:2009au,DelDuca:2010zg,Friot:2009fw,Friot:2011ic,Kalmykov:2012rr,Kalmykov:2016lxx,Loebbert:2019vcj,Ananthanarayan:2020ncn,Ananthanarayan:2020xpd}. Recently, MB integrals and the Mellin transform entered the conformal bootstrap, see \textit{e.g.} \cite{Gopakumar:2016wkt,Sleight:2019hfp} and references therein.
Other recent and diverse applications exist as for instance in option pricing \cite{Aguilar}, detector physics \cite{Friot:2014ufa} or RKKY interaction in condensed matter \cite{Oriekhov}.

Even though MB integrals have been thoroughly studied for several decades in theoretical physics, and in fact for more than one century in the mathematical literature - from the pionneering works \cite{Pincherle,Mellin,Barnes} to the most recent advances (see \textit{e.g.} \cite{NPT} and references therein) - it has been recently emphasized in \cite{Kalmykov_talk, Kalmykov:2020cqz} that there is still no systematic computational technique for the extraction of their multiple series representations in the $N$-fold case when $N>2$ (for the $N=2$ case with straight contours see \cite{Passare:1996db,TZ,Friot:2011ic}).

We present here the first solution to this important problem which, in addition to its own interest in the theory of MB integrals, can potentially lead to many new results in the fields mentioned above. A \textit{Mathematica} implementation of our method is given in the Supplemental Material to this paper, along with important specific examples of application of our method. The code is used, among others, to obtain the first evaluation of two highly nontrivial resonant cases in QFT: the hexagon and double box conformal Feynman integrals with unit propagator powers (see \cite{Ananthanarayan:2020ncn} for the nonresonant generic propagator powers cases).

\medskip

\indent{\textit{The method---}}  
The type of series representations that can be derived from Eq.(\ref{N_MB}) strongly depends on the $N$-dimensional vector ${\bf\Delta}=\sum_i a_i {\bf e}_i-\sum_j b_j {\bf f}_j$.  If ${\bf\Delta}$ is null, which is the case we focus on in the present work, this corresponds to a degenerate situation \cite{Passare:1996db, TZ} where there exist several convergent series representations for the MB integral, converging in different regions of the ${\bf x}$ parameter space. These series are analytic continuations of one another if the quantity $\alpha\doteq\text{Min}_{\vert\vert{\bf y}\vert\vert=1}(\sum_i a_i \vert{\bf e}_i\cdot{\bf y}\vert-\sum_jb_j \vert{\bf f}_j\cdot{\bf y}\vert)$ is positive \cite{Passare:1996db}. 

The question, now, is how to derive these series representations.
To ease the reading of the presentation of our method, which rests on a simple geometric analysis, we focus here on the nonresonant case where there is no point in the ${\bf z}$-space at which more than $N$ singular (hyper)planes (associated with the gamma functions in the numerator of the integrand of the $N$-fold MB integral) intersect. The poles of the MB integrand are thus of order one, thereby avoiding a discussion on the technical aspects of multivariate residues computations because only nonlogarithmic series representations can appear. Resonant, \textit{i.e} logarithmic cases, are discussed in the Supplemental Material, as well as in \cite{Ananthanarayan:2020xpd}.

To illustrate the different steps of the method, we propose to consider the simple paradigmatical example of the Appell $F_1$ double hypergeometric function whose MB representation reads \cite{KdF}:
 \begin{align}\label{2_MB_Appell_generic}
& F_{1}(a,b_1,b_2;c;u_1,u_2) = \frac{\Gamma(c)}{\Gamma(a)\Gamma(b_1)\Gamma(b_2)}
\nonumber \\
& \times \int_{-i \infty}^{+i \infty} \frac{d z_{1}}{2 \pi i}  \int_{-i \infty}^{+i \infty} \frac{d z_{2}}{2 \pi i} (-u_1)^{z_{1}} (-u_2)^{z_{2}} 
\Gamma(-z_1)\Gamma(-z_2) \nonumber \\ & \times \frac{\Gamma\left(a+z_{1}+z_{2}\right) \Gamma\left(b_1+z_{1}\right) \Gamma\left(b_2+z_{2}\right)}{\Gamma\left(c+z_{1}+z_{2}\right)}  
\end{align}
where the contours of integration are such that they separate the sets of poles of $\Gamma(-z_1)$ and $\Gamma(-z_2)$ from those of the other gamma functions in the numerator of the MB integrand. To avoid resonant situations, we choose generic values for the parameters $a,b_1,b_2$ and $c$.
It can be seen from Eq.(\ref{2_MB_Appell_generic}) that ${\bf\Delta}=(0,0)$ which means that this is a degenerate case, and a simple analysis shows that $\alpha=2$. Therefore, as mentioned above, one can conclude that the different series representations of the twofold MB integral that we will derive are analytic continuations of one another, converging in different regions of the $(u_1,u_2)$ space.

In the general MB case, each of the series representations that we look for is a particular linear combination of some multiple series. In the nonresonant case, such a linear combination is obtained as a sum of terms suitably extracted from a set $S$ of what we call \textit{building blocks} in the following. The latter are thus nothing but the multiple series dressed with their overall coefficient and sign. 

The key-point of our method (in the nonresonant case) is that each of these building blocks is associated with one $N$-combination of gamma functions in the numerator of the MB integrand and with one conic hull, and that specific intersections of these conic hulls are in one-to-one correspondence with the sums of building blocks which form the different series representations of the MB integral under study (in the resonant case, the same intersections give birth to series representations which are however not made of building blocks).

Let us see this in more details.
For each possible $N$-combination of gamma functions in the numerator of the MB integrand, let us consider the pointed conic hull, built from the vectors ${\bf e}_i$ of the gamma functions which belong to the $N$-combination. 
An $N$-combination whose associated conic hull is $N$-dimensional is retained, while the $N$-combinations yielding lower dimensional objects are discarded. Finding all relevant $N$-combinations, one therefore obtains a set of corresponding conic hulls, that we call $S'$, where $\text{Card}(S')=\text{Card}(S)$.

To see this in our $F_1$ example, let us label each of the five gamma functions of the integrand's numerator of Eq.\eqref{2_MB_Appell_generic} by $i=1,\cdots,5$ to keep track of them, and display them in a tabular form (see TABLE \ref{table_1}) along with their corresponding normal vector ${\bf e}_i$ and what we call their singular factor $s_i({\bf z})$, defined in Eq.(\ref{N_MB}).
\begin{table}[h] \label{Appell_Table}
\begin{center}
 \begin{tabular}{||c c c c||}
 \hline
 $i$ & $\Gamma$ function & $\mathbf{e}_{i}$ & $s_i({\bf z})$ \\ [0.5ex] 
 \hline\hline
 1 & $\Gamma(-z_1)$ & $(-1,0)$ & $-z_1$ \\ 
 2 & $\Gamma(-z_2)$ & $(0,-1)$ & $-z_2$ \\
 3 & $\Gamma(a+z_1+z_2)$ & $(1,1)$ & $z_1+z_2$ \\
 4 & $\Gamma(b_1+z_1)$ & $(1,0)$ & $z_1$ \\
 5 & $\Gamma(b_2+z_2)$ & $(0,1)$ & $z_2$ \\[0.5ex] 
 \hline
 \end{tabular}
 \end{center}
\caption{List of gamma functions in the numerator of the integrand in Eq.\eqref{2_MB_Appell_generic} and their associated normal vectors and singular factors.\label{table_1}}
\end{table}
Now, since the MB integral is twofold, one has to consider all possible 2-combinations $(i_1,i_2)$ of these gamma functions and their associated conic hulls $C_{i_1,i_2}$, where $i_1$ and $i_2$ are the labels, given in the first column of TABLE \ref{table_1},  of the gamma functions that belong to a given 2-combination. There are $\binom{5}{2}=10$ possible 2-combinations, out of which only eight are retained as for the two 2-combinations $(1,4)$ and $(2,5)$ the associated conic hulls are of lower dimension than the fold of the MB integral. 

\begin{figure}[h]
       \centering
       \includegraphics[width=2.8cm]{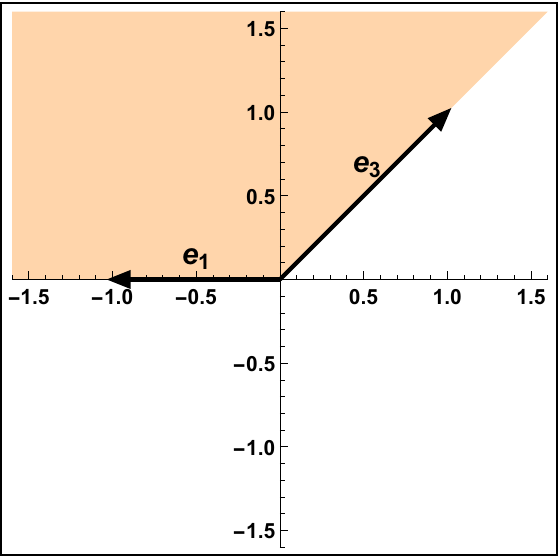}
       \includegraphics[width=2.8cm]{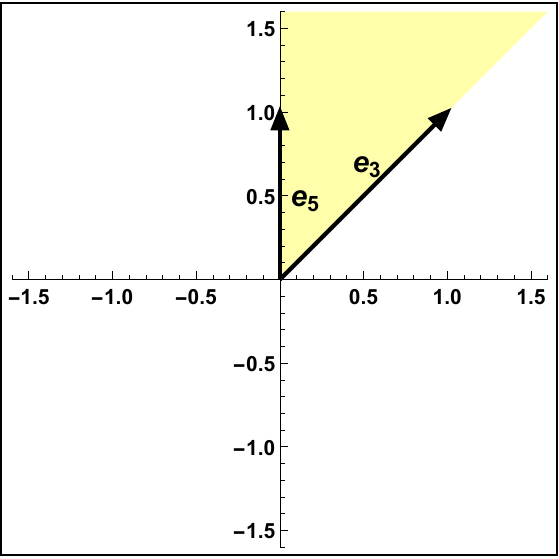}
       \includegraphics[width=2.8cm]{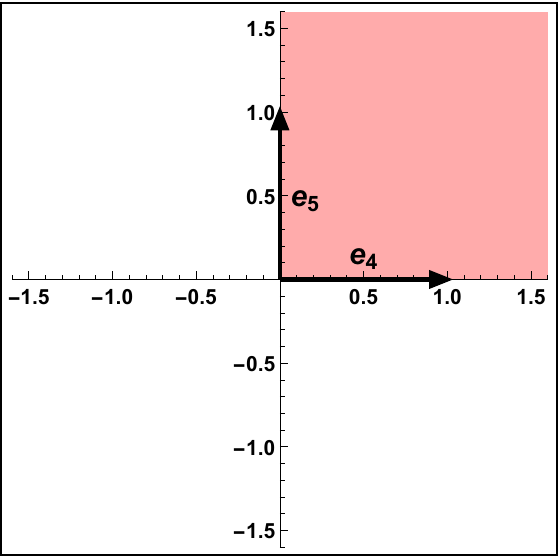}
       \includegraphics[width=2.8cm,valign=t]{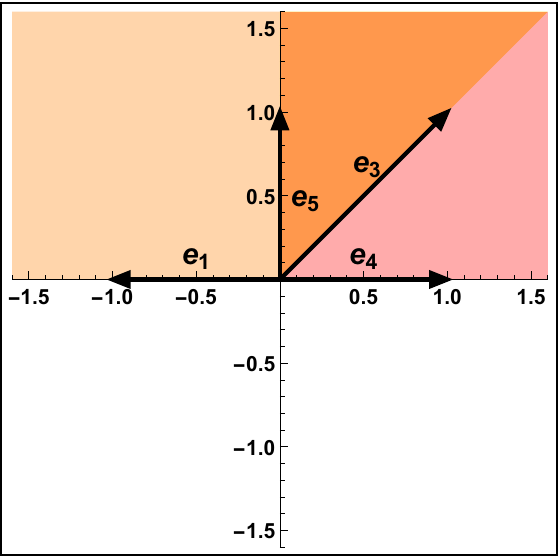}
        \includegraphics[width=3.2cm,valign=t]{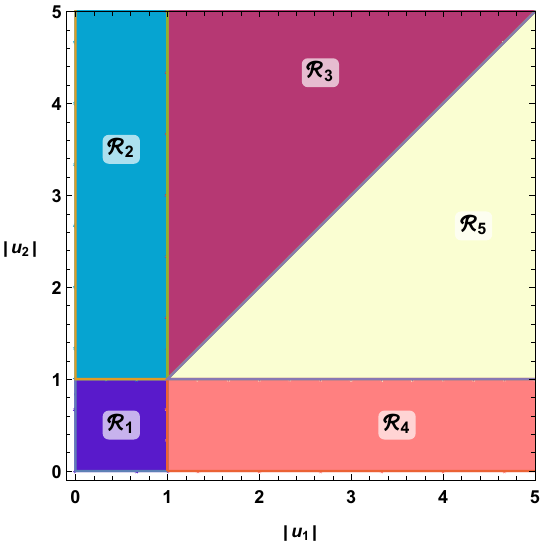}
       \caption{Conic hulls $C_{1,3}$ (top-left), $C_{3,5}$ (top-middle) and $C_{4,5}$ (top-right) and their intersection (orange region, bottom-left) with edges along  $\mathbf{e}_{3}$ and $\mathbf{e}_{5}$ which corresponds in fact to $C_{3,5}$. Bottom-right: Convergence regions, given in Eq.(\ref{Appell_Series_Representation}), of the five series representations of the MB integral in Eq.\eqref{2_MB_Appell_generic}. The well-known Appell $F_1$ double hypergeometric series converges in $\mathcal{R}_1$, its 4 analytic continuations in the other regions.}
       \label{Appell_13}
   \end{figure}

This way, the set of conic hulls associated with the retained 2-combinations is
   \begin{equation}
        S'=\big\{ C_{1,2}\,, C_{1,3} \,, C_{1,5}\,, C_{2,3}\,, C_{2,4}\,, C_{3,4}\,, C_{3,5}\,, C_{4,5}\big\}
   \end{equation}
As an example, the conic hull $C_{1,3}$ associated with $(1,3)$, whose edges are along the vectors $\mathbf{e}_{1}=(-1,0)$ and $\mathbf{e}_{3}=(1,1)$, is shown in FIG.\ref{Appell_13} (top-left). $C_{3,5}$ (resp. $C_{4,5}$) is shown in the top-middle (resp. top-right).
   \\
   
As mentioned above, one can now associate with each retained 2-combination $(i_1,i_2)$ a \textit{building block}, denoted by $B_{i_1,i_2}$. Consequently, $S$ simply reads
   \begin{equation}
       S=\big\{ B_{1,2}\,, B_{1,3} \,, B_{1,5}\,, B_{2,3}\,, B_{2,4}\,, B_{3,4}\,, B_{3,5}\,, B_{4,5}\big\}\label{BBlist}
   \end{equation}
We now have to compute explicitly the expressions of each of these building blocks and find the series representations that can be built from them. Note that it is of course possible to perform these two steps in reverse order because our method does not rest on the convergence properties of the involved multiple series.

In the general case, to each retained $N$-combination, there is a corresponding set of poles located at the intersections of exactly $N$ singular (hyper)planes (those of the gamma functions in the $N$-combination) which, by a straightforward residue calculation, gives the corresponding building block in $S$. 
Following \cite{Friot:2011ic}, one begins by bringing the singularity to the origin using appropriate changes of variables on the MB integrand and one applies the generalized reflection formula $\Gamma(z-n)=\frac{\Gamma(1+z)\Gamma(1-z)(-1)^n}{z\ \Gamma(n+1-z)}$ $(n\in\mathbb{Z})$, on each of the singular gamma functions so that their singular part appears explicitly. 
It then remains, in order to get the residue, to divide the obtained expression by $\vert \text{det}A\vert$, where  $A=(A_{rs})_{1\leq r\leq N, 1\leq s\leq N}$ with $A_{rs}=({\bold e}_{i_r})_s$, to remove the $N$ singular factors in the denominator and to put the $z_i$, $i=(1,\cdots,N)$ to zero. Summing over all residues one then obtains the expression of the desired building block.

Let us show how this works by considering, for instance,  in Eq.(\ref{BBlist}), the case of $B_{1,3}$ which is the sum of residues of the poles associated with $(1,3)$, located at $(n_1,-a-n_1-n_2)$ for $n_i \in \mathbb{N}$ $(i=1,2)$.

One first brings the singularity to the origin using the changes of variable $z_1\rightarrow z_1+n_1$ and  $z_2\rightarrow z_2-a-n_1-n_2$. Then, applying the reflection formula on the singular gamma functions, the MB integrand becomes
 \begin{align}
    & \left(-u_1\right)^{z_1+n_1}\left(-u_2\right)^{z_2-a-n_1-n_2}
    \frac{ \Gamma \left(1-z_1\right) \Gamma \left(1+z_1\right)(-1)^{n_1}}{(-z_1)\Gamma \left(n_1+1+z_1\right)}  \nonumber \\
    &\times \frac{\Gamma(-z_2+a+n_1+n_2) \Gamma \left(1+z_1+z_2\right) \Gamma \left(1-z_1-z_2\right)}{(z_1+z_2)\Gamma\left(n_2+1-z_1-z_2\right)}
    \nonumber \\
    & \times  (-1)^{n_2}
    \frac{\Gamma(b_1+z_1+n_1)\Gamma(b_2+z_2-a-n_1-n_2)}{\Gamma(c+z_1+n_1+z_2-a-n_1-n_2)} \nonumber
   \end{align}
Now, since $\vert\text{det}A\vert$, where $A=
\begin{pmatrix}
-1 & 0 \\
1 & 1 
\end{pmatrix}$, gives 1, it remains to remove the singular factors $s_1=-z_1$ and $s_3=z_1+z_2$ from the denominator and to put $z_1=z_2=0$. Multiplying by the overall prefactor (ratio of gamma functions in Eq.(\ref{2_MB_Appell_generic})) and summing over $n_1$ and $n_2$ one then obtains the expression of the building block
   \begin{align}
     B_{1,3}&= \frac{\Gamma(c)}{\Gamma(a)\Gamma(b_1)\Gamma(b_2)}(-u_2)^{-a} \sum_{n_1,n_2 =0}^{\infty}\left(-\frac{u_1}{u_2}\right)^{n_1}\left(\frac{1}{u_2}\right)^{n_2} \nonumber\\
     &\times \frac{ \Gamma \left(a+n_1+n_2\right) \Gamma
   \left({b_1}+n_1\right) \Gamma \left(-a+{b_2}-n_1-n_2\right)}{\Gamma \left(n_1+1\right) \Gamma \left(n_2+1\right) \Gamma
   \left(-a+c-n_2\right)}\nonumber\\
   & =\frac{\Gamma(c)\Gamma(b_2-a)}{\Gamma(b_2)\Gamma(c-a)} (-u_2)^{-a}  \nonumber\\
   &\ \ \ \ \times F_1\left(a,b_1,a-c+1;a-b_2+1;\frac{u_1}{u_2},\frac{1}{u_2}\right)
   \end{align}
A similar analysis yields
 \begin{align}
    &B_{1,5} = \frac{\Gamma(c)\Gamma(a-b_2)}{\Gamma(a)\Gamma(c-b_2)}(-u_2)^{-b_2} \nonumber\\ 
     &\hspace{1cm} \times G_2\left(b_1,b_2,b_2-c+1;a-b_2;-u_1,-\frac{1}{u_2}\right)
   \end{align}
where $G_2$ is one of the Horn double hypergeometric series \cite{Srivastava}. It is thus straightforward, from similar calculations, to derive the explicit form of each of the building blocks of Eq.(\ref{BBlist}).

Let us now explain how to build the various series representations of the $N$-fold MB integral without any convergence analysis, which is among the significant features of this paper. We observe that there is a one-to-one correspondence between these series representations and the subsets of conic hulls of $S'$ whose intersection is nonempty, with the important constraint that if a subset of conic hulls satisfying the nonempty intersection condition is included in a bigger subset that also satisfies it, then the former does not correspond to a series representation. 
In order to write down the expression of the series representation associated with a given subset, one simply has to add the building blocks in $S$ that correspond to each of the conic hulls of the subset. Every subset of conic hulls in $S'$ satisfying the nonempty intersection condition will then lead to one distinct series representation of the MB integral.

  In the case of Eq.(\ref{2_MB_Appell_generic}), a straightforward geometrical analysis yields five subsets, which therefore leads to five series representations that are analytic continuations of one another. 
The subsets are $\{C_{1,2}\}, \{C_{1,3},C_{1,5}\}, \{C_{1,3},C_{3,5},C_{4,5}\}, \{C_{2,3},C_{2,4}\}$ and $\{C_{2,3},C_{3,4},C_{4,5}\}$. As an example, we have shown the intersection corresponding to the third subset in Fig.\ref{Appell_13} (bottom-left).

One therefore obtains
   \begin{align}\label{Appell_Series_Representation}
 & F_1(a;b_1,b_2,c;u_1,u_2) \nonumber \\
 & =\left\{\begin{array}{ll}
B^*_{1,2} & \text { for } |u_1|<1  \cap |u_2|<1  \hspace{0.5cm} (\text{$\mathcal{R}_1$})\\
B_{1,3}+B^*_{1,5} & \text { for } |u_1|<1  \cap \left\vert\frac{1}{u_2}\right\vert<1 \hspace{0.5cm} (\text{$\mathcal{R}_2$})\\
B_{1,3}+B^*_{3,5}+B_{4,5} & \text { for} \left\vert\frac{1}{u_1}\right\vert<1  \cap \left\vert\frac{u_1}{u_2}\right\vert<1 \hspace{0.5cm} (\text{$\mathcal{R}_3$})\\
B_{2,3}+B^*_{2,4} & \text { for} \left\vert\frac{1}{u_1}\right\vert<1  \cap |u_2|<1  \hspace{0.5cm} (\text{$\mathcal{R}_4$})\\
B_{2,3}+B^*_{3,4}+B_{4,5} & \text { for} \left\vert\frac{u_2}{u_1}\right\vert<1  \cap \left\vert\frac{1}{u_2}\right\vert<1 \hspace{0.5cm} (\text{$\mathcal{R}_5$})
\end{array}\right.
\end{align}
where the $^*$ on a building block indicates the master series (see below) associated with that series representation.
The series representation $B_{1,3}+B_{1,5}$ and $B_{2,3}+B_{2,4}$ coincide with Eq.(17) of \cite{Olsson} while $B_{1,3}+B_{3,5}+B_{4,5}$ and $B_{2,3}+B_{3,4}+B_{4,5} $ match with Eq.(22) of the same reference.

Obviously the  last two series representations of Eq.(\ref{Appell_Series_Representation}) could be deduced from the second and third ones by using the permutation symmetry  $F_1(a,b_1,b_2;c;u_1,u_2)=F_1(a,b_2,b_1;c;u_2,u_1))$.

\medskip

\indent{\textit{Master series---}} Until here, we did not discuss convergence issues, because our method does not need to solve for the latter in order to extract the different series representations from the MB integral. However, once obtained, one may need to know the convergence regions of the series. We will see now that by introducing \textit{master series}, this task can be greatly simplified.

In the degenerate case, the convergence region of a particular series representation of the MB integral 
is given by the intersection of the convergence regions of each of the
series of which the series representation is built. Therefore, one way to find
the convergence region of a series representation is to find the convergence region of each of these terms. 
Beyond triple or even double series, these convergence issues can be difficult open problems. Moreover, the higher $k$ and/or $N$ in Eq.(\ref{N_MB}) are, the more the linear combinations that constitute the series representations each have a large number of terms with different convergence properties. This also increases the complexity of the convergence analysis.

The alternative strategy that we propose is to find a set of poles that can parameterize,
up to a change of variables, all the poles associated with the considered series representation. We call this set the {\it master set}. One can then construct from the master set a single series, which we name
the {\it master series}, and we conjecture that its
convergence region will either coincide or be a subset of the convergence region of the series representation under consideration. In the former case, which happens when there is no gamma function in the denominator of the MB integrand (or when there is at most a finite number of cancellations of poles by the gamma functions in the denominator), this considerably simplifies the task to that of finding the region of convergence of only this series (this is the case for our $F_1$ example), while in the latter case, although not explicitly giving the convergence region of the series representation, this is of precious help to facilitate the numerical checks. Note that even when the convergence region of the master series is too complicated to be derived, it is of great utility because it is sufficient to find a single set of numerical values that make it converge, to have the whole series representation also converging for the same set of values (this point is clearly illustrated in the study of the resonant double box and hexagon Feynman integrals performed in the Supplemental Material).

In the case of higher-fold MB integrals, it is not straightforward to find the master set algebraically. 
We therefore propose a simpler technique, where we infer the master series from the $N$-dimensional conic hull 
(the \textit{master conic hull}) formed by the intersection of the conic hulls associated with the $N$-combinations from which the series representation
is built. First, one obtains the $N$ basis vectors ${\bf e}_i$ $(i=1,\cdots,N)$ of the master conic hull. Then the set of poles resulting from  the meeting of the singular (hyper)planes associated with the gamma functions $\Gamma({\bf e}_i \cdot z_i)$, $(i=1,\cdots,N)$ gives the master set. 
Although the direction of the basis vectors ${\bf e}_i$ is given, their magnitude has to be fixed in such a way that the master set 
parameterizes all the poles that correspond to the series representation, up to a change of variable. 
Note that it can happen, in some cases, that the master series built from the master set is in fact one of the building blocks. 
This is the case for our $F_1$ example above and it is illustrated in FIG.\ref{Appell_13} (bottom-left) for the third series of Eq.(\ref{Appell_Series_Representation}) where it is indeed clear that the plotted intersection is a conic hull which matches with $C_{3,5}$. This means that $B_{3,5}$ is the master series associated with the third series representation of Eq.(\ref{Appell_Series_Representation}). Therefore, the convergence region of $B_{3,5}$ coincides with the region $\mathcal{R}_3$ (this can be easily checked by explicitly computing the intersection of the convergence regions of $B_{1,3}$, $B_{3,5}$ and $B_{4,5}$). In FIG. \ref{Appell_13} (bottom-right),  we show the convergence regions obtained from a study of the master series, indicated by a star in Eq.(\ref{Appell_Series_Representation}), of each series representation of Eq.(\ref{2_MB_Appell_generic}).

   We close this section by noting that, as far as the master series is concerned, resonant and nonresonant situations are treated in the same way.

\medskip

\indent{\textit{Conclusions---}} A new, and so far unique, simple and powerful systematic method for deriving series representations of $N$-fold MB integrals has been presented. It has the great advantage of selecting the different terms that form these series representations without the need of a prior study of the convergence regions of each of these terms. In the degenerate case, for each of the so obtained series representations, our method also allows one, in general, to derive a single master series. We have shown how the latter considerably simplifies the convergence analysis and/or the numerical checks.

We have also shown that our method can be used to deal with resonant (\textit{i.e} logarithmic) situations in the Supplemental Material as well as in our recent work \cite{Ananthanarayan:2020xpd}. In the latter paper, in addition to show an interesting interplay between QFT and hypergeometric functions theory, our method has been used to identify spurious contributions of a recent Yangian bootstrap approach used to compute Feynman integrals \cite{Loebbert:2020hxk}.

To show that investigations in cases with a high number of variables are not an unrealistic goal using our framework, we have applied it to ninefold MB integrals in  \cite{Ananthanarayan:2020ncn}, obtaining recently for the first time some series representations of the hexagon and double box conformal Feynman integrals, for generic powers of the propagators. Although these objects are very complicated, earlier attempts to compute them having failed (see for instance \cite{Loebbert:2019vcj}), they were easily computable with our approach because their MB representations belong to the nonresonant class. This is due to the fact that the propagator powers of these Feynman integrals are generic. Note that it is generally advised to compute Feynman integrals for generic powers of the propagators with the MB technique (see \cite{Smirnov:2012gma}). The same is true for multiple hypergeometric functions which are in general studied for generic values of their parameters. This gives us one more reason to believe that the efficiency and simplicity of our approach in the nonresonant case will give birth to many new results.

All the examples mentioned until here belong to the so-called degenerate class, where ${\bf \Delta}=0$, but our method can also treat the ${\bf \Delta}\neq0$ case where diverging asymptotic series representations can be obtained, as it will be shown in a subsequent publication.

We finish here by mentioning that we have provided, in the Supplemental Material, the first version of a \textit{Mathematica} implementation of our method. It gave, in less
than two minutes of CPU time, a series representation consisting of 26 terms for the hexagon \cite{Ananthanarayan:2020ncn}. In contrast
the \texttt{MBsums} \textit{Mathematica} package of \cite{Ochman:2015fho} gives a hardy usable linear combination of 112368 terms in over 12 hours on the same computer. We have also used our code to derive the first series representations of the hexagon and double box in the highly nontrivial resonant case of unit propagator powers.

\medskip

\begin{center}
\textbf{SUPPLEMENTAL MATERIAL}
\end{center}

\appendix
\
This Supplemental Material is composed of two appendices.
In Appendix A, the \textit{Mathematica} implementation of our computational method of multiple Mellin-Barnes integrals is presented, with some examples of application. Note that the code goes beyond the nonresonant class discussed at length in the main core of the paper because, as we shall see below, it can deal with resonant cases as well. 
These particular situations require a few more intermediate computation steps than the nonresonant ones. Therefore, we will explain what these steps are in Appendix B. We will also consider particular hybrid situations which have a nonresonant form at the end of the calculations, but ask for the computational technique of the resonant case for their evaluation. 

\section{Computer implementation}

{\it Basic usage--} The method described in this paper has been automated as a \textit{Mathematica} computer package called  \textsc{MBConicHulls} which has been written and tested in v12.2 of \textit{Mathematica} and does not run in versions lower than v12.0. It calls upon functions from the \textsc{MultivariateResidues} package \cite{Larsen:2017aqb} which has to be pre-installed as a dependency. Before describing the features of its functions, we mention that one of the tests passed by our package was the derivation of the analytic continuation formulas (59)-(69) presented in Chapter 9 of \cite{Srivastava}. All have been checked, except the trivial Gauss hypergeometric analytic continuation Eq. (60), since the package, in its present form, does not run for 1-fold MB integrals. Eq. (65) nor could not be derived because it involves the Horn $H_2$ function which does not have a simple MB representation.

Four functions need to be called by the user in order to evaluate any MB integral with our package:
\begin{itemize}
    \item {\tt MBRep[PreFac,IntVar,MBVar,MBArg]}: inputs the MB integral in a form that can be processed by the package.
    \begin{itemize}
        \item {\tt PreFac} is the prefactor of the integral.
        \item {\tt IntVar} takes in the list of integration variables $z_i$ of Eq.(1).
        \item {\tt MBVar} takes in the list of parameters $x_i$ of Eq.(1), each of which is raised to one $z_i$ in turn.
        \item {\tt MBArg} is the list consisting of two sublists of the form 
        {\tt \{\{numerator\},\{denominator\}\}}, where the elements of {\tt \{numerator\}} are the $s_i({\bf z})+g_i$ and the elements of {\tt \{denonimator\}} are the $t_j({\bf z})+h_j$ of Eq.(1).
    \end{itemize}
    
    \item {\tt ResolveMB[MBRepOut,N]}: returns the type of integral (degenerate or nondegenerate), the total number of associated conic hulls and then goes on to display the sets of poles, the master series characteristic list, and variables, for each series.
    \begin{itemize}
        \item {\tt MBRepOut} is the output of the {\tt MBRep} function that takes in the MB integral.
        \item {\tt N} is an optional parameter indicating the total number of series representations of the integral that one wishes to extract. If {\tt N}  is not specified, or if the value given by the user for {\tt N} is bigger than the number of possible series representations, then all series are shown. 
    \end{itemize}
    
    \item {\tt EvaluateSeries[ResolveMBOut, MBParaSub, SeriesNum]}:  calculates and returns the explicit expression of the series representation {\tt SeriesNum}.

    \begin{itemize}
        \item {\tt ResolveMBOut} is the output of the {\tt ResolveMB} function.
        \item {\tt MBParaSub} is a list of substitutions to be made to the parameters in the arguments of the gamma functions of the integrand (if none, just put empty braces). As particular substitution values may change the singular structure which is analyzed by {\tt EvaluateSeries[]}, it is important to substitute them at this stage or at the first step  {\tt MBRep[]}.
                \item {\tt SeriesNum} is the number, as enumerated in the output of {\tt ResolveMB[MBRepOut,N]}, of the series for which we wish to calculate the analytic expression.
    \end{itemize}
    
    \item {\tt SumAllSeries[EvaluateSeriesOut,MBVarSub,\\SumLim,
    RunInParallel,NumericalPrecision]}: numerically sums the particular series representation derived by {\tt EvaluateSeries}.
    \begin{itemize}
        \item {\tt EvaluateSeriesOut} is the output of the {\tt EvaluateSeries} function, consisting of the analytic expression of the selected series.
        \item {\tt MBVarSub} is a list of substitutions to provide numerical values to the $x_i$.
        \item {\tt SumLim} is the upper limit of the summation variables of the series representation.
         \item {\tt RunInParallel -> Bool} is an optional parameter indicating whether the sum is to be performed using Mathematica's parallel processing functionality. The default value of {\tt Bool} is {\tt False}.
         \item {\tt NumericalPrecision ->PositiveIntegers} is an optional parameter which determines the precision of the numerical sum. The default value is {\tt MachinePrecision}.
    \end{itemize}
\end{itemize}

We now demonstrate the use of this code on several examples, beginning with the Appell $F_1$ case used in the main core of the paper to illustrate our method. The implementation of these examples can be found in the \textit{Mathematica} notebook {\tt Examples.nb} accompanying this package.\\

{\it Example 1: Appell $F_1$--}  We first load the package (assuming that the \textsc{MBConicHulls} and \textsc{MultivariateResidues} packages and the notebook are in the same directory):\\

\textsf{{\scriptsize In[1]}}\textsf{{\scriptsize:{\tt=}}} {\tt SetDirectory[NotebookDirectory[]];}

\textsf{{\scriptsize In[2]}}\textsf{{\scriptsize:{\tt=}}} {\tt <<MBConicHulls.wl;} 

{\tt B.Ananthanarayan, S.Banik, S.Friot, S.Ghosh }\\

There is no need to explicitly call the \textsc{MultivariateResidues} package, which is directly called by our package internally.

One then inputs Eq.(2) as follows: \\

 \textsf{{\scriptsize In[3]}}\textsf{{\scriptsize:{\tt=}}} {\tt MBRepOut = MBRep[Gamma[c]/(Gamma[a]Gamma[${\mathtt b_{\mathtt1}}$] Gamma[${\mathtt b_{\mathtt2}}$]), \{${\mathtt z_{\mathtt1}}$, ${\mathtt z_{\mathtt2}}$\}, \{-${\mathtt u_{\mathtt1}}$, -${\mathtt u_{\mathtt2}}$\}, \{\{-${\mathtt z_{\mathtt1}}$, -${\mathtt z_{\mathtt2}}$, a + ${\mathtt z_{\mathtt1}}$ + ${\mathtt z_{\mathtt2}}$, ${\mathtt b_{\mathtt1}}$ + ${\mathtt z_{\mathtt1}}$, ${\mathtt b_{\mathtt2}}$ +${\mathtt z_{\mathtt2}}$\}, \{c + ${\mathtt z_{\mathtt1}}$ + ${\mathtt z_{\mathtt2}}$\}\}];}\\

\noindent Then one proceeds with the finding of the different features of the MB integral and, \textit{e.g}, of the first two series representations by calling:\\

 \textsf{{\scriptsize In[4]}}\textsf{{\scriptsize:{\tt=}}}{\tt ResolveMBOut = ResolveMB[MBRepOut, 2];} \\

\noindent which prints the output:\\

\noindent
{\tt Degenerate case with 8 conic hulls} \\

\noindent {\tt Series Solution 1::Intersecting Conic Hulls \{${\mathtt C_{{\mathtt1},{\mathtt2}}}$\}. The set of poles are :: \{\{${\mathtt n}_{\mathtt1},{\mathtt n}_{\mathtt2}$\}\} with master series characteristic list and variables \{\{${\mathtt n}_{\mathtt1},{\mathtt n}_{\mathtt2}$\},\{$-{\tt u}_{\tt2},-{\tt u}_{\tt1}$\}\}} \\

\noindent {\tt Series Solution 2::Intersecting Conic Hulls \{${\tt C}_{{\tt1},{\tt3}}$,${\tt C}_{{\tt1},{\tt5}}$\}. The set of poles are :: \\ \{\{${\tt n}_{\tt1},-{\tt a}{\tt-n}_{\tt1}{\tt-n}_{\tt2}$\},\{${\tt n}_{\tt1},{\tt-b}_{\tt2}{\tt-n}_{\tt2}$\}\} with master series characteristic list and variables  \{\{${\mathtt n}_{\mathtt1},{\mathtt n}_{\mathtt2}$\},\{${\tt-1}/{\tt u}_{\tt2},-{\tt u}_{\tt1}$\}\}} \\

\noindent {\tt Time Taken 0.201008 seconds} \\

\noindent The indicated time is for a 1.2 GHz Intel i7 4-core 2020 Apple MacBook Air with 16 GB of memory. 

Let us now obtain the expression of the second of these series representations for some chosen values of $a, b_1, b_2$ and $c$:\\

 \textsf{{\scriptsize In[5]}}\textsf{{\scriptsize:{\tt=}}} {\tt EvaluateSeriesOut= EvaluateSeries[\\ResolveMBOut, \{ a -> 1, ${\tt b}_{\tt1}$ -> 1/2, ${\tt b}_{\tt2}$ -> 1/3, c -> 1/4 \}, 2];}\\ \\
which prints the explicit residue series:\\

\noindent {\tt The series solution is a sum of the following 2 series.}\\

\noindent
 {\tt Series Number 1 :: \\
$\frac{ (-1)^{n_1+n_2} \Gamma(\frac{1}{4}) \Gamma(\frac{1}{2}+n_1)\Gamma(-\frac{2}{3}-n_1-n_2) \Gamma(1+n_1+n_2) (-u_1)^{n_1} (-u_2)^{-1-n_1-n_2}}{\sqrt{\pi}\Gamma(\frac{1}{3})\Gamma(-\frac{3}{4}-n_2)\Gamma(1+n_1)\Gamma(1+n_2)}$\\

 valid for $n_1 \geq 0 \, {\tt\&\&} \, n_2 \geq 0$}\\

\noindent
 {\tt Series Number 2 ::\\ 
$ \frac{ (-1)^{n_1+n_2}\Gamma(\frac{1}{4}) \Gamma(\frac{1}{2}+n_1)\Gamma(\frac{2}{3}+n_1-n_2) \Gamma(\frac{1}{3}+n_2) (-u_1)^{n_1} (-u_2)^{-\frac{1}{3}-n_2}}{\sqrt{\pi}\Gamma(\frac{1}{3})\Gamma(-\frac{1}{12}+n_1-n_2)\Gamma(1+n_1)\Gamma(1+n_2)}$\\

valid for $n_1 \geq 0 \, \&\& \, n_2 \geq 0$ \\

 Time Taken 2.9633 seconds
}\\

\noindent A straightforward calculation shows that the two above contributions correspond to Eqs.(5) and (6) in the main core of the paper, with $a=1$, $b_1=\frac{1}{2}$, $b_2=\frac{1}{3}$ and $c=\frac{1}{4}$. To sum these numerically for $0 \leq n_1 , n_2 \leq 15 $ and, for example, $u_1 = -0.3$ and $u_2 = -10.1$, we call\\

\textsf{{\scriptsize In[6]}}\textsf{{\scriptsize:{\tt=}}} {\tt SumAllSeries[EvaluateSeriesOut, \{${\tt u}_{\tt1}$->-0.3, ${\tt u}_{\tt2}$->-10.1\}, 15]}\\ 

\noindent {\tt Numerical Result: -0.212049}\\
\noindent {\tt Time Taken 0.244757 seconds}\\

\noindent and we see that the output matches the result obtained from the \textit{Mathematica}'s inbuilt Appell $F_1$ function call:\\

\textsf{{\scriptsize In[7]}}\textsf{{\scriptsize:{\tt=}}} {\tt AppellF1[${\tt1}$, $\frac{{\tt1}}{{\tt2}}$, $\frac{{\tt1}}{{\tt3}}$, $\frac{{\tt1}}{{\tt4}}$, -0.3,-10.1]}.\\

\textsf{{\scriptsize Out[7]}}\textsf{{\scriptsize:{\tt=}}} {\tt -0.212049}
 \\

{\it Other examples: the Hexagon and Double Box conformal Feynman integrals with unit propagator powers--}
These highly nontrivial resonant cases can be handled with our code, but the resulting expressions are too lengthy to be given here. Two particular series representations, built from respectively 64 and 140 series of nine variables, are however calculated and explicitly shown in the provided {\tt Examples.nb} notebook. The corresponding master series allowed us to easily find values of the nine variables that could be used to check the numerical matching between these series representations and the Feynman parameterizations of the hexagon and double box. A second check has been to numerically verify that the differential equation that links the hexagon to the double box is satisfied. The latter check has been performed at a 71 decimal places level, far below the size of the smallest series of each series representation, which guarantees that the contributions of all series of the series representations derived in the notebook have been tested. 
These examples provide another nontrivial test of the \textsc{MBConicHulls} package.

Note that although in principle the code can give all of the many possible series representations of the hexagon and double box, its present version cannot do it in a decent time. Further improvement of the code will aim to solve this computational time issue.

\section{}
We have shown in the main manuscript how efficient and simple our conic hulls technique is, in the nonresonant (\textit{i.e} nonlogarithmic) case which happens for generic values of the MB parameters.
In this appendix, we will be interested in the resonant case, which requires a few more intermediate computation steps than the nonresonant one. We will also consider particular hybrid situations which have a nonresonant form at the end of the calculations, but which ask for the computational technique of the resonant case in order to be evaluated.

To illustrate the two interesting situations mentioned above, we propose to re-examine the simple Appell $F_1$ case, used as an example in the main manuscript, for two different sets of values of $a,b_1,b_2$ and $c$. The first set, where $a=2, b_1=1, b_2=1$ and $c=1/2$ will generate a resonant case, whereas the second, where $a=2, b_1=1/2, b_2=1/2$ and $c=1$, will correspond to an hybrid situation.

In the general resonant situation, the poles of the MB integrand of the $N$-fold MB integral, coming from the intersections of more than $N$ singular (hyper)planes, are of higher multiplicity than in the nonresonant case. Therefore, the residue computations are more tricky but, as in the nonresonant situation, our method starts by finding all the relevant $N$-combinations\footnote{The possible powers of the involved gamma functions  are ignored for the determination of the relevant $N$-combinations.} of gamma functions in the numerator of the MB integrand, as well as the largest subsets of conic hulls in $S'$ having nonempty intersections. Let us call $S''$ the set of $N$-combinations that correspond to one of the obtained subsets of conic hulls. What differentiates the resonant case from the nonresonant one is that for a given $N$-combination in $S''$, parts or all of its associated set of poles can also be poles associated with some of the other $N$-combinations of $S''$, betraying the presence of poles of higher and possibly different multiplicities. Therefore the simple analysis, based on building blocks, that we have presented for the nonresonant case in the main manuscript, is no longer valid to find the series representations, in general\footnote{Note that it can happen that parts of the sets of singularities of a resonant case are of order one. For these simple poles one therefore has to follow the nonresonant approach previously described.}. One instead has to consider the $N$-combinations of $S''$ and, for each of them, one has to determine the different types of associated singularities, carefully avoiding possible double countings from one $N$-combination to another.

Once this has been performed, one has, for each type of singularities located at the intersections of more than $N$ singular (hyper)planes, to divide the set of singular factors of the related gamma functions into $N$ suitable groups $f_i({\bf z})$ $(i=1, \cdots,N)$, for the need of the multivariate residues computation (more precisely for the transformation law \cite{Larsen:2017aqb}). Note that it may be difficult, or even perhaps impossible, to build such a single set of $N$ groups that we denote as the vector ${\bf f}({\bf z})\doteq(f_1({\bf z}),f_2({\bf z}),\cdots,f_N({\bf z}))$. In this case one has to deal with a suitable sum $\sum_\alpha{\bf f_\alpha}({\bf z})$ of such vectors which will give equivalent although less compact results at the end (see Eq.(\ref{Appell_grouping_2}) for an example with such a sum). Note that these vectors have to be zero-dimensional ideals, which means that for ${\bf f}({\bf z})$ to satisfy this property, the solution of $f_1({\bf z})=f_2({\bf z})=\cdots=f_N({\bf z})=0$ has to consist of a finite number of points ${\bf z}$.

For a given type of singularities associated with one of the relevant $N$-combinations, the derivation of the vector(s) of $N$ groups of singular factors, as well as their corresponding residues, proceeds as follows.

As in the nonresonant case one begins by bringing the singularity to the origin and by applying the generalized reflection formula on each of the singular gamma functions. Suitable combinations of their singular factors  will form each of the $N$ groups $f_i({\bf z})$ $(i=1, \cdots,N)$. If the $i^{th}$ gamma function $\Gamma^{a_i}(s_i ({\bf z})+g_i)$ in the numerator of the MB integrand of Eq.(1) of the main manuscript is singular at the considered type of poles, its singular factor will have the form $s_i ({\bf z})=({\bf e}_{i}\cdot{\bf z})^{a_{i}}$. We then list   
the singular factors of all the gamma functions in the numerator of the MB integrand that are singular at the poles under consideration, in a set $G$.  
In fact, each $N$-combination of gamma functions that belongs to $S''$ will contribute to the calculation of the residues associated with the type of singularities under consideration if its singular factors form a subset of $G$ (in the case where several $N$-combinations contribute, double counting has to be avoided by considering a given type of singularities only once).

Now we consider the singular factors associated with each of the $N$-combinations that contribute and show how to deduce the (combination of) vector(s) grouping these singular factors.
For this we write the contribution of one of the involved $N$-combinations as $(s_{i_1}({\bf z}),..., s_{i_N}({\bf z}))$ where, as said above, the $s_{i_j}({\bf z})$ $(j=1, \cdots,N)$ belong to $G$. 
Let us define the following rules (we now remove the ${\bf z}$ dependency of the $s_i({\bf z})$ to lighten the equations)
\begin{align}
(s_{i_1},\cdots,s_{i_{k-1}}, s_{i_k},s_{i_{k+1}}&,\cdots,s_{i_N})\nonumber\\
+&(s_{i_1},\cdots,s_{i_{k-1}}, s'_{i_k},s_{i_{k+1}},\cdots,s_{i_N}) \nonumber\\
=(s_{i_1},\cdots,s_{i_{k-1}}, s_{i_k}s'_{i_k}&,s_{i_{k+1}},\cdots,s_{i_N})\label{rule1}
\end{align}
and
\begin{align}
 &(s_{i_1},\cdots, s_{i_{k-1}}, s_{i_k},s_{i_{k+1}},s_{i_{k+2}},\cdots,s_{i_N})=-(s_{i_1},\cdots\nonumber\\
&\hspace{3cm}, s_{i_{k-1}}, s_{i_{k+1}},s_{i_{k}},s_{i_{k+2}},\cdots,s_{i_N})\label{rule2}
\end{align}

The aim is now, starting from the RHS of the following formal equation (\ref{f_1-f_N}), to derive the simplest form of its LHS using rules (\ref{rule1}) and (\ref{rule2}):
\begin{equation}\label{f_1-f_N}
   \sum_\alpha{\bf f_\alpha}({\bf z})\ =\tilde\sum S_{i_1,i_2,\cdots,i_N}(s_{i_1},\cdots,s_{i_N})
\end{equation}
where $S_ {i_1, \cdots ,i_N} \doteq \text{sign}( \text{det} A)$
and
$A=(A_{rs})_{1\leq r\leq N, 1\leq s\leq N}\hspace{0.4cm}\text{with}\hspace{0.4cm}A_{rs}=({\bold e}_{i_r})_s$.

As explained above, we stress that the $\tilde\sum$ sum sign in the RHS of Eq.\eqref{f_1-f_N} is over all the $N$-combinations that correspond to the series representation under consideration and whose singular factors form a subset of $G$. And the $\sum_\alpha$ sum sign in the LHS recalls that it can happen that the result is obtained as a combination of sets of $N$ groups of singular factors, instead of a single one.
In this case, each vector ${\bf f}_\alpha({\bf z})$ is subject to the condition that it contains contributions of all the singular factors in $G$ and that it is a zero-dimensional ideal.

One must note here that Eq.\eqref{f_1-f_N} may not have a unique solution (see Eqs.(\ref{Appell_grouping_1}) and (\ref{Appell_grouping_2}) for an example of such situation). However, from our experience, any solution that satisfies Eq.\eqref{f_1-f_N} will give the same result.

It may also happen that some gamma functions in the denominator of the MB integrand are also singular at the considered type of singularity. In this case one applies the procedure for the explicit extraction of their singular factors (which this time appear in the numerator) and one simplifies the final form of the grouping accordingly. We show examples where this simplification has to be taken into account in \cite{Ananthanarayan:2020xpd} and in Eq.(\ref{deno_simpl}) below.

The residues of the considered type of singularities are then obtained by adding the residues corresponding to each vector ${\bf f}_\alpha({\bf z})$. For the explicit computation of these residues one has to perform the transformation law \cite{griffiths}. Using the ${\bf f}_\alpha({\bf z})$ as inputs, this can be done automatically with the help of the $\textsc{MultivariateResidues}$ package. This step is performed by our code which calls this package internally.

After the computation of the contribution of this particular type of singularities, it is necessary to look for other types of singularities associated with the same $N$-combination, if any. Once done, one has to go on with the next $N$-combination in $S''$ and perform the same analysis (avoiding double counting). The final answer for the series representation is obtained by adding the contributions of all $N$-combinations in $S''$.\\

Let us see how this works in the resonant $F_1$ example
\begin{align}\label{2_MB_Appell_resonant}
& F_{1}(2,1,1;1/2;u_1,u_2) =\nonumber \\
&  \Gamma(1/2) \int_{-i \infty}^{+i \infty} \frac{d z_{1}}{2 \pi i}  \int_{-i \infty}^{+i \infty} \frac{d z_{2}}{2 \pi i} (-u_1)^{z_{1}} (-u_2)^{z_{2}} 
\Gamma(-z_1)\Gamma(-z_2) \nonumber \\
& 
 \nonumber \\ & \times \frac{\Gamma\left(2+z_{1}+z_{2}\right) \Gamma\left(1+z_{1}\right) \Gamma\left(1+z_{2}\right)}{\Gamma\left(1/2+z_{1}+z_{2}\right)}  
\end{align}
Since the conic hulls depend only on the coefficient vectors $\mathbf{e}_i$, the relevant set of conic hulls is obviously the same as in the nonresonant case considered in the main manuscript. Therefore, the set $S'$ containing the largest subsets of conic hulls whose intersection is nonempty will also be the same.

Let us focus on the subset $\{C_{1,3}, C_{1,5} \}$.

We have $S''=\{(1,3),(1,5)\}$ and the associated poles are at $(n_1,-2-n_1-n_2)$ and $(n_1,-1-n_2)$, which we now consider individually. 
\medskip

$\bullet$ Set 1: Poles at $(n_1,-2-n_1-n_2)$ associated with the 2-combination $(1,3)$.
    
   We shift the poles to the origin by substituting $z_1 \to z_1+n_1$ and $z_2 \to z_2-2-n_1-n_2$ in the integrand of Eq.\eqref{2_MB_Appell_resonant},
    \begin{align} \label{Appell_2_Set1}
       & \sqrt{\pi}(-u_1)^{z_{1}+n_1} (-u_2)^{z_{2}-2-n_1-n_2} 
\Gamma(2+n_1+n_2-z_2) \nonumber \\
&  \times \frac{\Gamma(-z_1-n_1)\Gamma\left(-n_2+z_{1}+z_{2}\right) }{\Gamma\left(-3/2-n_2+z_{1}+z_{2}\right)}
\nonumber \\
& \times \Gamma\left(1+n_1+z_{1}\right) \Gamma\left(-1-n_1-n_2+z_{2}\right)
    \end{align}
    It is evident from the above expression that the second, third and fifth gamma functions in the numerator are singular at the origin, for all values of $n_i \in \mathbb{N}$. The number of these singular gamma functions being greater than the number of folds of the integral indicates that this is a resonant case. We can also infer that these poles will overlap with other sets of poles associated with combinations in $S''$. Since we have not considered any such poles yet we have not needed to worry about double counting.
    
       As in the nonresonant case we then apply the generalized reflection formula on each singular gamma function in Eq.\eqref{Appell_2_Set1}, to obtain
    \begin{align}
       & -\sqrt{\pi}(-u_1)^{z_{1}+n_{1}} (-u_2)^{z_{2}-2-n_1-n_2} \Gamma(1-z_1)\Gamma(1+z_1)
 \nonumber \\
&  \times \frac{\Gamma(-z_2+2+n_1+n_2)\Gamma(1-z_1-z_2)\Gamma(1+z_1+z_2)}{(-z_1)(z_1+z_2)z_2\Gamma(1+z_1+n_1)\Gamma\left(1+n_2-z_{1}-z_{2}\right)} \nonumber \\
&  \times  \frac{\Gamma(z_1+1+n_1)\Gamma(1+z_2)\Gamma(1-z_2)\Gamma(2+n_1-z_1-z_2)}{\Gamma(2+n_1+n_2-z_2)\Gamma(-3/2-n_2+z_1+z_2)}
    \end{align}
    and the set $G$ of singular factors is $\{s_1, s_3 , s_5 \}=\{-z_1,z_1+z_2,z_2\}$.
    
    We next group these three denominator singular factors in $G$ as $(f_1,f_2)$,  using Eq.(\ref{f_1-f_N}). Each $N$-combination $(i_1,i_2)$ on the RHS of Eq.(\ref{f_1-f_N})  must satisfy the conditions that it must be an element of the set $S''$ and its singular factors $\{s_{i_1},s_{i_2}\}$ must belong to the set $G$.  Therefore, we have
    \begin{align}
    \label{Appell_grouping_2_Set_1}
        (f_1 , f_2) & = S_{1,3} (s_1, s_3) + S_{1,5} (s_1, s_5) = -(s_1, s_3)-(s_1, s_5) \nonumber \\ & = -(s_1, s_3s_5) = ((z_1+z_2)z_2, -z_1)
    \end{align}
    where we have used the rules in Eq.(\ref{rule1}) and Eq.(\ref{rule2}).

    Finally, using the \textsc{MultivariateResidues} package with $(f_1,f_2)$ in Eq.\eqref{Appell_grouping_2_Set_1} we obtain the logarithmic result:
    \begin{align} \label{Appell_2_1}
        S_1 = -\sqrt{\pi} \sum_{n_1,n_2=0}^{\infty} & \frac{ \log (-u_2) - \psi(-3/2-n_2) + \psi (1+n_2)}{\Gamma(-3/2-n_2)\Gamma(1+n_2)} \nonumber \\ & \times \left(\frac{u_1}{u_2}\right)^{n_1} \left(- \frac{1}{u_2}\right)^{n_2+2} \nonumber\\
        = -\frac{\sqrt{\pi}}{u_1-u_2} \sum_{n_2=0}^{\infty} & \frac{ \log (-u_2) - \psi(-3/2-n_2) + \psi (1+n_2)}{\Gamma(-3/2-n_2)\Gamma(1+n_2)} \nonumber \\ & \hspace{1.5cm} \times  \left(- \frac{1}{u_2}\right)^{n_2+1}
    \end{align}

$\bullet$ Set 2: Poles at $(n_1,-1-n_2)$ of $(1,5)$.
    
    Shifting the poles to the origin one gets
    \begin{align} 
       & \sqrt{\pi} (-u_1)^{z_1+n_1} (-u_2)^{z_2-1-n_2} \Gamma(-z_1-n_1)
 \nonumber \\
&  \times \frac{\Gamma(-z_2+1+n_2)\Gamma\left(1+n_1-n_2+z_{1}+z_{2}\right)}{\Gamma\left(-1/2+n_1-n_2+z_{1}+z_{2}\right)} \nonumber \\
&  \times  \Gamma\left(1+n_1+z_{1}\right) \Gamma\left(-n_2+z_{2}\right)
    \end{align}
    where one sees that $n_1$ and $n_2$ do not have the same sign in one of the singular gamma functions. We thus have to consider two different possible situations: $n_2 \leq n_1$ and $n_2 \geq n_1+1$.
    
    First, let us consider the case $n_2 \leq n_1$. Here only the first and fifth numerator gamma functions are singular. This is a nonresonant case as the number of singular gamma functions is equal to the number of folds of the MB integral. Therefore, a simple analysis leads to the following contribution:
    \begin{align} \label{Appell_2_2}
        S_2 &= -\frac{\sqrt{\pi}}{u_2}\sum_{n_1=0}^{\infty} \sum_{n_2=0}^{n_1} \frac{\Gamma(1+n_1-n_2)}{\Gamma(-1/2+n_1-n_2)} (u_1)^{n_1} \left(\frac{1}{u_2}\right)^{n_2} \nonumber\\
        &= -\frac{\sqrt{\pi}}{u_2}\sum_{n_1,n_2=0}^{\infty} \frac{\Gamma(1+n_1)}{\Gamma(-1/2+n_1)} (u_1)^{n_1} \left(\frac{u_1}{u_2}\right)^{n_2}
        \nonumber\\
        &= \frac{_2F_1(1,1;-\frac{1}{2};u_1)}{2(u_2-u_1)}
    \end{align}
    where $_2F_1(a,b;c;x)$ is the Gauss hypergeometric series.\\
    Let us now consider the case $n_2 \geq n_1+1$. Here three gamma functions of the numerator are singular, indicating that the poles are overlapping with another set of poles. A straightforward analysis confirms that this overlap is with Set 1 which has already been considered. Therefore, we omit the contribution of this case to avoid double counting.

    Hence, the sum of the series in Eq.\eqref{Appell_2_1} and Eq.\eqref{Appell_2_2} gives the series representation of the MB integral (\ref{2_MB_Appell_resonant}) for the subset $\{C_{1,3}, C_{1,5}\}$ which is valid for $|u_1| <1 \cap |u_2| >1$ (the convergence region being the same as that of the associated master series). Taking for instance the same values $u_1=-0.3$ and $u_2=-10.1$ as before, one can once again check the numerical agreement with \textit{Mathematica}'s inbuilt Appell $F_1$ function.
    \medskip

As mentioned in the beginning of this section, we now want to consider an hybrid situation where the resonant approach is needed at an intermediate step in the calculations although the expressions obtained at the end of the calculations have a non-resonant form (\textit{i.e} it is nonlogarithmic). This example will also give us an explicit realization of our statements about the grouping of the singular factor which can in some case have several equivalent forms, the latter being possibly splitted in several vectors of $N$ groups. 

For this, once again we come back to our favorite Appell function, choosing $a=2$, $b_1=b_2=1/2$ and $c=1$:
\begin{align}\label{2_MB_Appell_resonantnonresonant}
& F_{1}(2,1/2,1/2;1;u_1,u_2) = 
\nonumber \\
& \frac{1}{\Gamma^2(1/2)} \int_{-i \infty}^{+i \infty} \frac{d z_{1}}{2 \pi i}  \int_{-i \infty}^{+i \infty} \frac{d z_{2}}{2 \pi i} (-u_1)^{z_{1}} (-u_2)^{z_{2}} 
\Gamma(-z_1)\Gamma(-z_2) \nonumber \\
& 
 \nonumber \\ & \times \frac{\Gamma\left(2+z_{1}+z_{2}\right) \Gamma\left(1/2+z_{1}\right) \Gamma\left(1/2+z_{2}\right)}{\Gamma\left(1+z_{1}+z_{2}\right)}  
\end{align}
As before the associated set of conic hulls and its relevant subsets are the same as those of the nonresonant case. 

Let us elaborate our discussion of the series representation corresponding to the subset of conic hulls $\{C_{1,3}, C_{3,5}, C_{4,5} \}$ with $S''=\{ (1,3),(3,5),(4,5)\}$ whose associated poles are at $(n_1,-2-n_1-n_2)$, $(-3/2-n_1+n_2,-1/2-n_2)$ and $(-1/2-n_1,-1/2-n_2)$. We will consider each type of pole separately.

$\bullet$ Set 1: Poles at $(n_1,-2-n_1-n_2)$ of $(1,3)$.
    
    Shifting the poles to the origin we get
    \begin{align}
       & (-u_1)^{z_{1}+n_1} (-u_2)^{z_{2}-2-n_1-n_2} 
\Gamma(-z_1-n_1) \nonumber \\
&  \times \frac{\Gamma(-z_2+2+n_1+n_2)\Gamma\left(-n_2+z_{1}+z_{2}\right) }{\pi \Gamma\left(-1-n_2+z_{1}+z_{2}\right)}
\nonumber \\
& \times \Gamma\left(1/2+n_1+z_{1}\right) \Gamma\left(-3/2-n_1-n_2+z_{2}\right)
    \end{align}
    It is obvious from the above expression that the first and third gamma functions in the numerator, as well as the denominator gamma function, are singular at the origin, for all values of $n_i \in \mathbb{N}$. Hence, the contribution from this set of poles is null as there are in fact no poles in $z_2$.
    
$\bullet$ Set 2: Poles at $(-3/2-n_1+n_2,-1/2-n_2)$ of $(3,5)$.
    
    This case gives
    \begin{align} \label{Appell_Set2}
       & (-u_1)^{z_{1}-3/2-n_1+n_2} (-u_2)^{z_{2}-1/2-n_2} 
 \nonumber \\
&  \times \frac{\Gamma(-z_1+3/2+n_1-n_2)\Gamma(-z_2+1/2+n_2)}{\pi \Gamma\left(-1-n_1+z_{1}+z_{2}\right)} \nonumber \\
&  \times \Gamma\left(-n_1+z_{1}+z_{2}\right) \Gamma\left(n_2-1-n_1+z_{1}\right) \Gamma\left(-n_2+z_{2}\right)
    \end{align}
    We observe that for $n_2 \geq 2+n_1$, the residue of the pole at the origin is zero as there are no poles in $z_1$. On the other hand, for $n_2 \leq 1+n_1$, the third, fourth and fifth gamma functions of the numerator, as well as the denominator gamma function, are singular, which will lead to nonzero residues.
    
  Strictly speaking this case is not logarithmic because the singular gamma function in the denominator will lower the order of the singularities coming from the numerator, giving birth to a nonresonant situation. However, one cannot avoid the use of the resonant formalism in the intermediate steps to compute the corresponding contributions.  
  
  One can predict that these poles will overlap with other sets of poles associated with other combinations in $S''$. But we have not considered any such poles as no combination containing the fourth gamma function has been evaluated so far.
    
    To calculate the residue we apply the generalized reflection formula on each singular factors in Eq.\eqref{Appell_Set2}, to get the simplified analytic part
    \begin{align}
       & (-1)^{n_1} (-u_1)^{z_{1}-3/2-n_1+n_2} (-u_2)^{z_{2}-1/2-n_2} 
 \nonumber \\
&  \times \frac{\Gamma(-z_1+3/2+n_1-n_2)\Gamma(-z_2+1/2+n_2)\Gamma(1+z_1)}{\pi \Gamma\left(1+n_1-z_{1}-z_{2}\right)} \nonumber \\
&  \times  \frac{\Gamma(1-z_1)\Gamma(1+z_2)\Gamma(1-z_2)\Gamma(2+n_1-z_1-z_2)}{\Gamma(2+n_1-n_2-z_1)\Gamma(1+n_2-z_2)}
    \end{align}
    and the set $G$ of singular factors is $\{s_3,s_4,s_5 \}=\{z_1+z_2,z_1,z_2 \}$ while there is a factor $(z_1+z_2)$ in the numerator.
    
    We next group the three denominator singular factors in $G$ as $(f_1,f_2)$,  using Eq.(\ref{f_1-f_N}), for the need of the calculation of the residues. We recall that each $2$-combinations $(i_1,i_2)$ on the RHS of Eq.(\ref{f_1-f_N}) must satisfy the conditions that it is an element of the set $S''$, and its singular factors $\{s_{i_1},s_{i_2}\}$ must belong to the set $G$.  Therefore, we have
    \begin{align}
    \label{Appell_grouping_1}
        (f_1 , f_2) & = S_{3,5} (s_3,s_5) + S_{4,5} (s_4,s_5)  =  (s_3,s_5) +  (s_4,s_5) \nonumber \\ & = (s_3s_4, s_5) = ((z_1+z_2)z_1, z_2)
    \end{align}
    where we have used the rules in Eq.(\ref{rule1}) and Eq.(\ref{rule2}) of the main text.

We emphasize that the expression of $(f_1,f_2)$ is not unique. Indeed, an alternative expression is
\begin{align}
\label{Appell_grouping_2}
        (f_1 , f_2) & = (s_3,s_5) +  (s_4,s_5) \nonumber \\ & = (s_3,s_5) + (s_3,s_4) - (s_3,s_4)+ (s_4,s_5) \nonumber \\ & = (s_3,s_5) + (s_3,s_4) + (s_4,s_3)+ (s_4,s_5) \nonumber \\ & = (s_3,s_4s_5)+ (s_4,s_3s_5)\nonumber \\ &= (z_1+z_2, z_1 z_2)+ (z_1, (z_1+z_2) z_2)
    \end{align}
which gives the same residue as Eq.\eqref{Appell_grouping_1}. This can be more readily seen in our case due to the singular factor $(z_1+z_2)$ in the numerator, which originates from the singular gamma function in the denominator and has to be accounted for. Indeed, one has
    \begin{align}\label{deno_simpl}
        \frac{z_1+z_2}{ ((z_1+z_2)z_1, z_2)} &=\frac{1}{(z_1, z_2)}\nonumber  \\ &=    \frac{z_1+z_2}{ ((z_1+z_2),z_1 z_2)}+ \frac{z_1+z_2}{ (z_1, (z_1+z_2)z_2)}
    \end{align}
   the penultimate term being zero.
   
    Finally, evaluating the residue for $(f_1,f_2)=(z_1,z_2)$ we obtain the contribution of Set 2 which reads 
        \begin{align} \label{Series_Representation_3}
        R &=\frac{(u_1)^{-3/2}(u_2)^{-1/2}}{\pi}\sum_{n_1=0}^\infty\sum_{n_2=0}^{1+n_1}\left(\frac{1}{u_1}\right)^{n_1}\left(\frac{u_1}{u_2}\right)^{n_2}   \nonumber \\ & \times \frac{\Gamma(2+n_1)\Gamma(3/2+n_1-n_2)\Gamma(1/2+n_2)}{\Gamma(1+n_1)\Gamma(1+n_2)\Gamma(2+n_1-n_2)}
\nonumber\\        
        &= \frac{1}{2} \frac{(u_2)^{-1/2}}{(u_1-1)^{3/2}}+ (u_1)^{-1/2}(u_2)^{-3/2} \sum_{n_1,n_2=0}^\infty \left(\frac{1}{u_1}\right)^{n_1} \nonumber \\ & \times \left(\frac{1}{u_2}\right)^{n_2} \frac{\Gamma(1/2+n_1)\Gamma(3/2+n_2)\Gamma(2+n_1+n_2)}{{\pi} \, n_1! \, (n_2+1)! \,\Gamma(1+n_1+n_2)}
    \end{align}
    where the second term in the last equality is a Kampé-de Fériet series converging in $|u_1| > 1 \cap |u_2|>1$.
    
$\bullet$ Set 3: Poles at $(-1/2-n_1, -1/2 -n_2)$ of $(4,5)$.
    
    On shifting the poles to the origin, we obtain
    \begin{align}
        & (-u_1)^{-1/2-n_1+z_1} (-u_2)^{-1/2-n_2+z_2} \Gamma(1/2+n_1-z_1) \nonumber \\ & \times  \frac{\Gamma(1/2+n_2-z_2) \Gamma(1-n_1-n_2+z_1+z_2)}{\pi \Gamma(-n_1-n_2+z_1+z_2)} 
        \nonumber \\ & \times\Gamma(-n_1+z_1)\Gamma(-n_2+z_2)
    \end{align}
    We note that for $n_1+n_2 < 1$, the residue is zero. For $n_1+n_2 \geq 1$, the third, fourth and fifth gamma functions and the denominator gamma function are singular. As there are more than two numerator gamma functions, this again indicates that this set of poles overlap with some other set of poles. A straightforward analysis confirms that this set of poles overlaps with the $n_2 \leq 1+n_1$ case of Set 2. Thus, to avoid double counting we discard these poles which have already been considered.

    Hence, Eq.\eqref{Series_Representation_3} gives the series solution of the MB integral for the subset $\{C_{1,3}, C_{3,5}, C_{4,5} \}$ which is valid for $|u_1| >1 \cap |u_2| >1$.
    
    Let us give a brief remark on the master series conjecture before ending our discussion. We recall that, for the above considered subset of conic hulls, the convergence region of the master series was found to be $|u_1|<|u_2| \cap |u_1| >1$ (see Eq.(7) in the main manuscript), which is smaller than the convergence region of the solution in Eq.\eqref{Series_Representation_3}. This is not unexpected as there were an infinite number of cancellations of poles due to the gamma function in the denominator. However, even in such cases one cannot undermine the role of master series as for higher-fold MB it can be extremely difficult to find the convergence region of the series representations. Therefore, the fact that the convergence region of the master series is a subset of that of the corresponding series representation makes it useful for numerical checks.

\medskip

\indent{\textit{Acknowledgments---}} We thank Alankar Dutta and Vijit Kanjilal for technical
assistance. S. G. thanks Collaborative Research Center CRC 110 Symmetries and the Emergence of Structure in QCD for supporting the research through grants.


\begin{thebibliography}{}




\bibitem{KdF}
  P.~Appell and J.~Kamp\'e de F\'eriet, 
  \textit{``Fonctions hyperg\'eom\'etriques et hypersph\'eriques - Polyn\^omes d'Hermite''},
  Gautiers-Villars et $\text{C}^{\text{ie}}$, 1926.

\bibitem{Exton}
  H.~Exton, 
  \textit{``Multiple hypergeometric functions and applications''},
 Ellis Horwood Series in Mathematics and Its Applications, 1976.

\bibitem{Marichev}
  O. I.~Marichev, 
  \textit{``Handbook of integral transforms of higher transcendental functions: Theory and Algorithmic tables''},
 Ellis Horwood Series in Mathematics and Its Applications, 1983.

\bibitem{Sasiela}
  R. J. Sasiela, 
  \textit{``Electromagnetic wave propagation in turbulence: evaluation and application of Mellin transforms''}, Springer Series on Wave Phenomena, Vol. 18, Springer-Verlag Berlin Heidelberg, 1994.


\bibitem{Paris&Kaminsky}
  R. B. Paris and D. Kaminski,\textit{``Asymptotics and Mellin-Barnes integrals"}, Encyclopedia of Mathematics and its Applications, Vol. 85, Cambridge University Press, Cambridge, 2001.

\bibitem{Smirnov:2012gma}
V.~A.~Smirnov,
\textit{``Analytic tools for Feynman integrals''},
Springer Tracts Mod. Phys. \textbf{250} (2012), 1-296
doi:10.1007/978-3-642-34886-0.


\bibitem{Bjorken:1963zz}
J.~D.~Bjorken and T.~T.~Wu,
Phys. Rev. \textbf{130} (1963), 2566-2572
doi:10.1103/PhysRev.130.2566

\bibitem{Trueman:1963zz}
T.~L.~Trueman and T.~Yao,
Phys. Rev. \textbf{132} (1963), 2741-2748
doi:10.1103/PhysRev.132.2741

\bibitem{Usyukina1975}
N. I. Usyukina,
Theor. Math. Phys. \textbf{22} (1975), 300-306.



\bibitem{Boos:1990rg}
E.~E.~Boos and A.~I.~Davydychev,
Theor. Math. Phys. \textbf{89} (1991), 1052-1063
doi:10.1007/BF01016805.

\bibitem{Davydychev:1990jt}
A.~I.~Davydychev,
J. Math. Phys. \textbf{32} (1991), 1052-1060
doi:10.1063/1.529383



\bibitem{Davydychev:1990cq}
A.~I.~Davydychev,
J. Math. Phys. \textbf{33} (1992), 358-369
doi:10.1063/1.529914

\bibitem{Usyukina:1992jd}
N.~I.~Usyukina and A.~I.~Davydychev,
Phys. Lett. B \textbf{298} (1993), 363-370
doi:10.1016/0370-2693(93)91834-A

\bibitem{Usyukina:1993ch}
N.~I.~Usyukina and A.~I.~Davydychev,
Phys. Lett. B \textbf{305} (1993), 136-143
doi:10.1016/0370-2693(93)91118-7

\bibitem{Davydychev:1992mt}
A.~I.~Davydychev and J.~B.~Tausk,
Nucl. Phys. B \textbf{397} (1993), 123-142
doi:10.1016/0550-3213(93)90338-P

\bibitem{Berends_1994}
Berends, F.A., Böhm, M., Buza, M. et al. 
Z. Phys. C - Particles and Fields 63, 227–234 (1994). https://doi.org/10.1007/BF01411014

\bibitem{Usyukina:1994iw}
N.~I.~Usyukina and A.~I.~Davydychev,
Phys. Lett. B \textbf{332} (1994), 159-167
doi:10.1016/0370-2693(94)90874-5
[arXiv:hep-ph/9402223 [hep-ph]].



\bibitem{Smirnov:1999gc}
V.~A.~Smirnov,
Phys. Lett. B \textbf{460} (1999), 397-404
doi:10.1016/S0370-2693(99)00777-7
[arXiv:hep-ph/9905323 [hep-ph]].

\bibitem{Tausk:1999vh}
J.~B.~Tausk,
Phys. Lett. B \textbf{469} (1999), 225-234
doi:10.1016/S0370-2693(99)01277-0
[arXiv:hep-ph/9909506 [hep-ph]].

\bibitem{Smirnov:2000vy}
V.~A.~Smirnov,
Phys. Lett. B \textbf{491} (2000), 130-136
doi:10.1016/S0370-2693(00)00997-7
[arXiv:hep-ph/0007032 [hep-ph]].

\bibitem{Smirnov:2001cm}
V.~A.~Smirnov,
Phys. Lett. B \textbf{524} (2002), 129-136
doi:10.1016/S0370-2693(01)01382-X
[arXiv:hep-ph/0111160 [hep-ph]].

\bibitem{Smirnov:2003vi}
V.~A.~Smirnov,
Phys. Lett. B \textbf{567} (2003), 193-199
doi:10.1016/S0370-2693(03)00895-5
[arXiv:hep-ph/0305142 [hep-ph]].

\bibitem{Heinrich:2004iq}
G.~Heinrich and V.~A.~Smirnov,
Phys. Lett. B \textbf{598} (2004), 55-66
doi:10.1016/j.physletb.2004.07.058
[arXiv:hep-ph/0406053 [hep-ph]].




\bibitem{Czakon:2005rk}
M.~Czakon,
Comput. Phys. Commun. \textbf{175} (2006), 559-571
doi:10.1016/j.cpc.2006.07.002
[arXiv:hep-ph/0511200 [hep-ph]].

\bibitem{Anastasiou:2005cb}
C.~Anastasiou and A.~Daleo,
JHEP \textbf{10} (2006), 031
doi:10.1088/1126-6708/2006/10/031
[arXiv:hep-ph/0511176 [hep-ph]].




\bibitem{Gluza:2007rt}
J.~Gluza, K.~Kajda and T.~Riemann,
Comput. Phys. Commun. \textbf{177} (2007), 879-893
doi:10.1016/j.cpc.2007.07.001
[arXiv:0704.2423 [hep-ph]].


\bibitem{Smirnov:2009up}
A.~V.~Smirnov and V.~A.~Smirnov,
Eur. Phys. J. C \textbf{62} (2009), 445-449
doi:10.1140/epjc/s10052-009-1039-6
[arXiv:0901.0386 [hep-ph]].


\bibitem{Ochman:2015fho}
M.~Ochman and T.~Riemann,
Acta Phys. Polon. B \textbf{46} (2015) no.11, 2117
doi:10.5506/APhysPolB.46.2117
[arXiv:1511.01323 [hep-ph]].



\bibitem{Friot:2005cu}
S.~Friot, D.~Greynat and E.~De Rafael,
Phys. Lett. B \textbf{628} (2005), 73-84
doi:10.1016/j.physletb.2005.08.126
[arXiv:hep-ph/0505038 [hep-ph]].

\bibitem{Aguilar:2008qj}
J.~P.~Aguilar, D.~Greynat and E.~De Rafael,
Phys. Rev. D \textbf{77} (2008), 093010
doi:10.1103/PhysRevD.77.093010
[arXiv:0802.2618 [hep-ph]].

\bibitem{Czakon:2007wk}
M.~Czakon, A.~Mitov and S.~Moch,
Nucl. Phys. B \textbf{798} (2008), 210-250
doi:10.1016/j.nuclphysb.2008.02.001
[arXiv:0707.4139 [hep-ph]].

\bibitem{Smirnov:2009fh}
A.~V.~Smirnov, V.~A.~Smirnov and M.~Steinhauser,
Phys. Rev. Lett. \textbf{104} (2010), 112002
doi:10.1103/PhysRevLett.104.112002
[arXiv:0911.4742 [hep-ph]].

\bibitem{Lee:2010cga}
R.~N.~Lee, A.~V.~Smirnov and V.~A.~Smirnov,
JHEP \textbf{04} (2010), 020
doi:10.1007/JHEP04(2010)020
[arXiv:1001.2887 [hep-ph]].


\bibitem{Greynat:2012ww}
D.~Greynat and E.~de Rafael,
JHEP \textbf{07} (2012), 020
doi:10.1007/JHEP07(2012)020
[arXiv:1204.3029 [hep-ph]].

\bibitem{deRafael:2014gxa}
E.~de Rafael,
Phys. Lett. B \textbf{736} (2014), 522-525
doi:10.1016/j.physletb.2014.08.003
[arXiv:1406.4671 [hep-lat]].


\bibitem{Ananthanarayan:2016pos}
B.~Ananthanarayan, J.~Bijnens, S.~Ghosh and A.~Hebbar,
Eur. Phys. J. A \textbf{52} (2016) no.12, 374
doi:10.1140/epja/i2016-16374-8
[arXiv:1608.02386 [hep-ph]].

\bibitem{Charles:2017snx}
J.~Charles, E.~de Rafael and D.~Greynat,
Phys. Rev. D \textbf{97} (2018) no.7, 076014
doi:10.1103/PhysRevD.97.076014
[arXiv:1712.02202 [hep-ph]].



\bibitem{Ananthanarayan:2017qmx}
B.~Ananthanarayan, J.~Bijnens, S.~Friot and S.~Ghosh,
Phys. Rev. D \textbf{97} (2018) no.9, 091502
doi:10.1103/PhysRevD.97.091502
[arXiv:1711.11328 [hep-ph]].

\bibitem{Ananthanarayan:2018irl}
B.~Ananthanarayan, J.~Bijnens, S.~Friot and S.~Ghosh,
Phys. Rev. D \textbf{97} (2018), 114004
doi:10.1103/PhysRevD.97.114004
[arXiv:1804.06072 [hep-ph]].

\bibitem{Ananthanarayan:2020acj}
B.~Ananthanarayan, S.~Friot and S.~Ghosh,
Phys. Rev. D \textbf{101} (2020) no.11, 116008
doi:10.1103/PhysRevD.101.116008
[arXiv:2003.12030 [hep-ph]].




\bibitem{Bern:2005iz}
Z.~Bern, L.~J.~Dixon and V.~A.~Smirnov,
Phys. Rev. D \textbf{72} (2005), 085001
doi:10.1103/PhysRevD.72.085001
[arXiv:hep-th/0505205 [hep-th]].


\bibitem{Bern:2006vw}
Z.~Bern, M.~Czakon, D.~A.~Kosower, R.~Roiban and V.~A.~Smirnov,
Phys. Rev. Lett. \textbf{97} (2006), 181601
doi:10.1103/PhysRevLett.97.181601
[arXiv:hep-th/0604074 [hep-th]].



\bibitem{Bern:2006ew}
Z.~Bern, M.~Czakon, L.~J.~Dixon, D.~A.~Kosower and V.~A.~Smirnov,
Phys. Rev. D \textbf{75} (2007), 085010
doi:10.1103/PhysRevD.75.085010
[arXiv:hep-th/0610248 [hep-th]].

\bibitem{Drummond:2006rz}
J.~M.~Drummond, J.~Henn, V.~A.~Smirnov and E.~Sokatchev,
JHEP \textbf{01} (2007), 064
doi:10.1088/1126-6708/2007/01/064
[arXiv:hep-th/0607160 [hep-th]].

\bibitem{Kalmykov:2008ofy}
M.~Y.~Kalmykov, V.~V.~Bytev, B.~A.~Kniehl, B.~F.~L.~Ward and S.~A.~Yost,
PoS \textbf{ACAT08} (2008), 125
doi:10.22323/1.070.0125
[arXiv:0901.4716 [hep-th]].

\bibitem{DelDuca:2009au}
V.~Del Duca, C.~Duhr and V.~A.~Smirnov,
JHEP \textbf{03} (2010), 099
doi:10.1007/JHEP03(2010)099
[arXiv:0911.5332 [hep-ph]].



\bibitem{DelDuca:2010zg}
V.~Del Duca, C.~Duhr and V.~A.~Smirnov,
JHEP \textbf{05} (2010), 084
doi:10.1007/JHEP05(2010)084
[arXiv:1003.1702 [hep-th]].




\bibitem{Friot:2009fw}
S.~Friot and D.~Greynat,
SIGMA \textbf{6} (2010), 079
doi:10.3842/SIGMA.2010.079
[arXiv:0907.5593 [hep-th]].

\bibitem{Friot:2011ic}
S.~Friot and D.~Greynat,
J. Math. Phys. \textbf{53} (2012), 023508
doi:10.1063/1.3679686
[arXiv:1107.0328 [math-ph]].

\bibitem{Kalmykov:2012rr}
M.~Y.~Kalmykov and B.~A.~Kniehl,
Phys. Lett. B \textbf{714} (2012), 103-109
doi:10.1016/j.physletb.2012.06.045
[arXiv:1205.1697 [hep-th]].

\bibitem{Kalmykov:2016lxx}
M.~Y.~Kalmykov and B.~A.~Kniehl,
JHEP \textbf{07} (2017), 031
doi:10.1007/JHEP07(2017)031
[arXiv:1612.06637 [hep-th]].

\bibitem{Loebbert:2019vcj}
F.~Loebbert, D.~M\"uller and H.~M\"unkler,
Phys. Rev. D \textbf{101} (2020) no.6, 066006
doi:10.1103/PhysRevD.101.066006
[arXiv:1912.05561 [hep-th]].


\bibitem{Ananthanarayan:2020ncn}
B.~Ananthanarayan, S.~Banik, S.~Friot and S.~Ghosh,
Phys. Rev. D \textbf{102} (2020) no.9, 091901
doi:10.1103/PhysRevD.102.091901
[arXiv:2007.08360 [hep-th]].

\bibitem{Ananthanarayan:2020xpd}
B.~Ananthanarayan, S.~Banik, S.~Friot and S.~Ghosh,
Phys. Rev. D \textbf{103} (2021) no.9, 096008
doi:10.1103/PhysRevD.103.096008
[arXiv:2012.15646 [hep-th]].

\bibitem{Gopakumar:2016wkt}
R.~Gopakumar, A.~Kaviraj, K.~Sen and A.~Sinha,
Phys. Rev. Lett. \textbf{118} (2017) no.8, 081601
doi:10.1103/PhysRevLett.118.081601
[arXiv:1609.00572 [hep-th]].

\bibitem{Sleight:2019hfp}
C.~Sleight and M.~Taronna,
JHEP \textbf{02} (2020), 098
doi:10.1007/JHEP02(2020)098
[arXiv:1907.01143 [hep-th]].

\bibitem{Aguilar}
J.-P. Aguilar and J. Korbel, 
Fractal Fract. (2018), 2-15; doi:10.3390/fractalfract2010015.

\bibitem{Friot:2014ufa}
S.~Friot,
Nucl. Instrum. Meth. A \textbf{773} (2015), 150-153
doi:10.1016/j.nima.2014.10.026
[arXiv:1410.3985 [nucl-ex]].

\bibitem{Oriekhov}
D. O. Oriekhov and V. P. Gusynin,
Phys. Rev. B 101, 235162.





\bibitem{Pincherle}
  S.~Pincherle, Atti R. Accademia Lincei, Rend. Cl. Sci. Fis. Mat. Nat. (Ser. 4) {\bf 4}, 694 and 792 (1888).

\bibitem{Mellin}
R. H. Mellin, Acta Soc. Sci. Fenn. {\bf 20} (7), 1 (1895).

\bibitem{Barnes}
E. W. Barnes, Messenger Math. {\bf 29} (2), 64 (1900).


\bibitem{NPT}
L. Nilsson, M. Passare and A. Tsikh, J. Sib. Fed. Univ. Math. Phys., Vol. 12, Issue 4, 509-529  (2019).







\bibitem{Kalmykov_talk}
M. Kalmykov, \textit{``Hypergeometric functions and Feynman diagrams"}, Talk given at the ``Antidifferentiation and the Calculation of Feynman Amplitudes" Conference, 4-9 October 2020, Zeuthen, Germany.

\bibitem{Kalmykov:2020cqz}
M.~Kalmykov, V.~Bytev, B.~A.~Kniehl, S.~O.~Moch, B.~F.~L.~Ward and S.~A.~Yost,

[arXiv:2012.14492 [hep-th]].





\bibitem{Passare:1996db}
M.~Passare, A.~K.~Tsikh and A.~A.~Cheshel,
Theor. Math. Phys. \textbf{109} (1996), 1544-1555
doi:10.1007/BF02073871
[arXiv:hep-th/9609215 [hep-th]].

\bibitem{TZ}
O. Zhdanov and A. Tsikh, Siberian Mathematical Journal {\bf 39}, 245 (1998).




\bibitem{SM}
See Supplemental Material at http://link.aps.org/supplemental/




\bibitem{TZ2}
O. Zhdanov and A. Tsikh, Dokl. Math. 57, 24 (1998).



 \bibitem{Srivastava}
H.~M.~Srivastava and P.~W.~Karlsson \textit{``Multiple gaussian hypergeometric series''}, Ellis Horwood Series in Mathematics and Its Applications, 1985.  
 

\bibitem{Olsson}
P. O. M. Olsson, J. Math. Phys. {\bf 7}, 702 (1966).














\bibitem{Loebbert:2020hxk}
F.~Loebbert, J.~Miczajka, D.~M\"uller and H.~M\"unkler,
Phys. Rev. Lett. \textbf{125} (2020) no.9, 091602
doi:10.1103/PhysRevLett.125.091602
[arXiv:2005.01735 [hep-th]].







\bibitem{Larsen:2017aqb}
K.~J.~Larsen and R.~Rietkerk,
Comput. Phys. Commun. \textbf{222} (2018), 250-262
doi:10.1016/j.cpc.2017.08.025
[arXiv:1701.01040 [hep-th]].

\bibitem{griffiths}
P. Griffiths and J. Harris, \textit{Principles of Algebraic Geometry}, John Wiley \& Son, 1978. ISBN 0-471-32792-1.





\end{thebibliography}
\end{document}